\documentclass[letterpaper]{JHEP3}
\usepackage{graphicx}
\usepackage{epsfig}

\title{Counting Pockets with World Lines in Eternal Inflation}
\author{Richard Easther, Eugene A. Lim \\ Department of Physics\\ Yale University\\ New Haven, CT 06511 USA\\ E-mail: \email{richard.easther@yale.edu}\\ E-mail: \email{eugene.lim@yale.edu}}
\author{Matthew R Martin\\ T-8, MS B285\\ Los Alamos National Laboratory\\ Los Alamos, NM 87545 USA\\ E-mail: \email{mrmartin@lanl.gov}}

\abstract{We consider the long standing puzzle of how to obtain meaningful probabilities in eternal inflation.  We demonstrate a new algorithm to compute the probability distribution of pocket universe types, given a multivacua inflationary potential. The computed probability distribution is finite and manifestly gauge-independent. We argue that in some scenarios this technique can be applied to disfavor some eternally inflating potentials.}

\begin{document}

\section{Introduction}
Inflation~\cite{Guth:1980zm,Linde:1981mu,Albrecht:1982wi} has celebrated many successes. In addition to dynamically solving the cosmological problems of flatness, homogeneity and overabundance of monopoles, it also provides a mechanism for the generation of primordial fluctuations, which seed the large scale structure of the universe~\cite{Starobinsky:1982ee,Bardeen:1983qw}.

One of the most intriguing predictions of standard inflationary theory is that the same fluctuations which give rise to structure may cause inflation to be future eternal~\cite{OriginalEternalInflation}. Eternal inflation suggests that long before observable scales left the horizon during inflation, the quantum fluctuations of the inflaton were of the same order as the classical evolution of the inflaton rolling down its potential. Thus there were regions in space where the inflaton fluctuated up the potential and inflation proceeded more rapidly. On the other hand, there were also regions where the inflaton rolled or fluctuated down its potential.  In this case, non-inflating regions, or pocket universes~\cite{Guth:2000ka}, formed and we must live in one such region.  The entire space of pocket universes and inflating regions is sometimes called a multiverse. If the inflationary potential has more than one vacuum, then different pocket universes may have different values for observable parameters such as the coupling constants of the standard model or the value of the cosmological constant~\cite{Linde:1993hgjf,Tegmark:2004qd}.  Even in the same pocket universe, widely separated regions may have different values for some parameters.

Since we only have measurements from inside a single pocket, we are unable to sample the distribution of pockets.  Hence we cannot hope to reconstruct the model of eternal inflation observationally. This is an extreme version of \emph{cosmic variance} which prevents us, even in principle, from determining the exact underlying model which generates the spectrum of perturbations in the Cosmic Microwave Background~\cite{White:1993jr}. However, as we will discuss in Section~(\ref{sec:ProbParts}), there is still a notion of how well a particular theory or model of eternal inflation matches to our cosmic variance limited observations.

In addition, we lack a first principles derivation of the process of eternal inflation.  Indeed Vachaspati~\cite{Vachaspati:2003de} has an argument for why this derivation may not be possible within semi-classical gravity and thus one has to look towards quantum gravity for such a calculation.  

On the other hand, eternal inflation may provide a dynamical mechanism for sampling the many vacua that exist in some frameworks, such as the string theory landscape~\cite{Kachru:2003aw,Douglas:2003um,Douglas:2004zu,Feng:2000if,Bousso:2000xa}.  This mechanism underpins anthropic arguments which might explain the value of the cosmological constant, however unsatisfactory these may be from a philosophical perspective~\cite{Susskind:2003kw}.  More importantly, if eternal inflation is an unavoidable consequence of inflation then computing its implication for the probability distribution of fundamental parameters will give us valuable insight into the viability of various inflationary models as we will see in the next section.

Thus our goal in this work is to understand the statistics of parameters which vary across the infinite number of pocket universes, given some eternal inflationary potential. We assume that eternal inflation happens and our starting point is the multiverse structure. Specifically, we are interested in the probability of finding an observer in a region of the universe who measures parameters $\alpha_i$, given some inflationary potential, $V(\psi_j)$. The potential may be a function of many fields with many vacua. This probability distribution is a product of several separate, and distinct components (similar to~\cite{Garriga:2005av}):
\begin{equation}
P\left(\textrm{obs}_{\alpha_i} | V(\psi_j) \right) = 
\sum_A P\left(A|V(\psi_j)\right) P\left(\alpha_i |A\right) P(\textrm{obs} | \alpha_i). \label{prodprob}
\end{equation}
where $P(A|V)$ is the probability distribution of pocket universes of vacuum type $A$ in the potential, $P(\alpha_i|A)$ is the probability distribution of the set of parameters $\alpha_i$ in a pocket universe of type $A$, and $P(\textrm{obs}|\alpha_i)$ is a probability distribution for the existence of observers given parameters $\alpha_i$.  In some situations it is difficult to avoid questions about what is an observer, but we argue that those issues can be contained in this last factor. The final probability distribution is then a sum over all pockets $A$. Notice that we have added a subscript $\alpha_i$ to the total distribution on the left hand side, to emphasize that we have \emph{not} summed over parameters $\alpha_i$. We elaborate more on each individual distribution in the next section.

In this paper we propose a new method for calculating the first of these three probability factors, $P(A|V)$. This calculation involves finding a gauge-invariant regulator for the countably infinite number of pocket universes. We start with some finite set of points and velocities on the initial, uniformly inflating region. We then follow the world line corresponding to each initial point and velocity until it has entered a non-inflating region.  Different world lines may end up in regions which thermalize in different vacua.  We use these world lines to sample the different pockets and calculate probabilities based on the number of world lines in each type of pocket universe. Since multiple lines will end up in the same pocket, our method requires us to remove these duplicates in order to avoid over-counting. This also removes sensitivity to the initial distribution, as we will show. Finally, we take the limit as the number of world lines tends to infinity.

This paper is organized as follows. In Section~(\ref{sec:ProbParts}) we elaborate on the meaning of the probabilities involved. In Section~(\ref{sec:volbased}), we review previous work. We then introduce and describe our method in Section~(\ref{sec:ObserverMethod}), and present a sample computation that demonstrates how it works in practice.  We conclude in Section~(\ref{SummarySection}).

\section{Probability Distributions and Extreme Cosmic Variance}\label{sec:ProbParts}

Given a set of pocket universes generated by some potential $V$, the total probability of an observer measuring some set of fundamental properties $\alpha_i$ is expressed by equation~(\ref{prodprob}). 

The first distribution $P(A|V)$ of equation~(\ref{prodprob}) is the focus of our paper. This describes the probability of randomly drawing a pocket universe of type $A$ from an infinite set of pocket universes generated by $V$.  We will postpone the discussion of its computation to Section~(\ref{sec:ObserverMethod}). This distribution is clearly a function of the \emph{global} features, in field space, of the generating potential $V$.

Next, let us consider the second probability distribution $P(\alpha_i|A)$. Since the parameters may be a function of the value of the inflaton which might be inhomogenous (e.g. akin to some quintessence models \cite{Sandvik:2001rv}), we allow them to vary across the pocket. As shown by reference~\cite{Vanchurin:1999iv}, we can regulate this distribution by considering the volume fraction of the parameter $\alpha_i$ in a sphere of size $R$ and then taking $R$ to infinity. We will discuss this in Section~\ref{sec:volbased}. In contrast to the first distribution, these distributions are a function of the \emph{local} features around each respective minimum, in field space, of the generating potential $V$.

The third distribution however is much more ambiguous and difficult to define than the other two, as we have to decide what ``observers'' are and what are the parameters that allow for their existence. In this paper we focus on the value of the cosmological constant. We simply ignore all anthropic arguments (akin to the ``bottom-up'' approach of~\cite{Aguirre:2005cj}) and set $P(\textrm{obs}_{\alpha_i}|\alpha_i)$ to a constant as follows 
\begin{equation}
P(\textrm{obs}_{\rho_{\Lambda}}|0 \leq \rho_{\Lambda}<\rho_{\textrm{crit}})\propto \textrm{constant},
\quad P(\textrm{obs}_{\rho_{\Lambda}}| \rho_{\Lambda}=\rho_{\textrm{crit}}) = 0, \label{lambdadist}
\end{equation}
where the second part encodes the uncontroversial assumption that a pure de Sitter space cannot support perturbations~\cite{Mukhanov:1990me} and therefore is uninteresting. 

With these definitions in hand, let us illustrate an application of equation~(\ref{prodprob}). Imagine that our potential $V$ has three types of vacua labeled by ``us'', ``0'' and ``$1$'', with corresponding minima at the values of $\rho_{\Lambda}^{\textrm{us}}=0.7\rho_{\textrm{crit}}$, $\rho_{\Lambda}^0=0$ and $\rho_{\Lambda}^{1}=\rho_{\textrm{crit}}$. Since the inflaton might not settle at the absolute minima of these vacua, we allow these ``constants'' to vary across their respective pocket universes; for simplicity we assume that the middle distribution for each is a gaussian with identical variance\footnote{Its value is not important to our discussion.} $\sigma$ but different means
\begin{equation}
P(\rho_{\Lambda}^A|A)\propto e^{-(\rho_{\Lambda}-\rho_{\Lambda}^A)^2/(\sigma\rho_{crit})^2}.
\end{equation} 
For definiteness, we take $P(\textrm{obs}|\rho_{\Lambda})$ to be given by distribution~(\ref{lambdadist}),

Now, assuming that eternal inflation will produce a distribution of pockets such that one of the above three types dominates, then there are three possibilities depending on which pocket is most common. We label the potentials that generate these possibilities $V_1$, $V_2$ and $V_3$. If $V_1$ generates the distributions such that $P(\textrm{us}|V_1)\gg P(0|V_1),~P(1|V_1)$, as in Figure~\ref{fig:totalprobabilitygood}, then we conclude that we are \emph{typical} observers of this universe since our single data point is $\rho_{\Lambda}=0.7\rho_{\textrm{crit}}$. On the other hand, if for potential $V_2$ we find, $P(0|V_2) \gg P(\textrm{us}|V_2),~P(1|V_2)$, as shown in Figure~\ref{fig:totalprobabilitybad}, then we are \emph{atypical} observers of this universe since our observed value of $\rho_{\Lambda}$ is at the tail end of the gaussian peak at the origin.

\FIGURE[t]{\epsfig{file=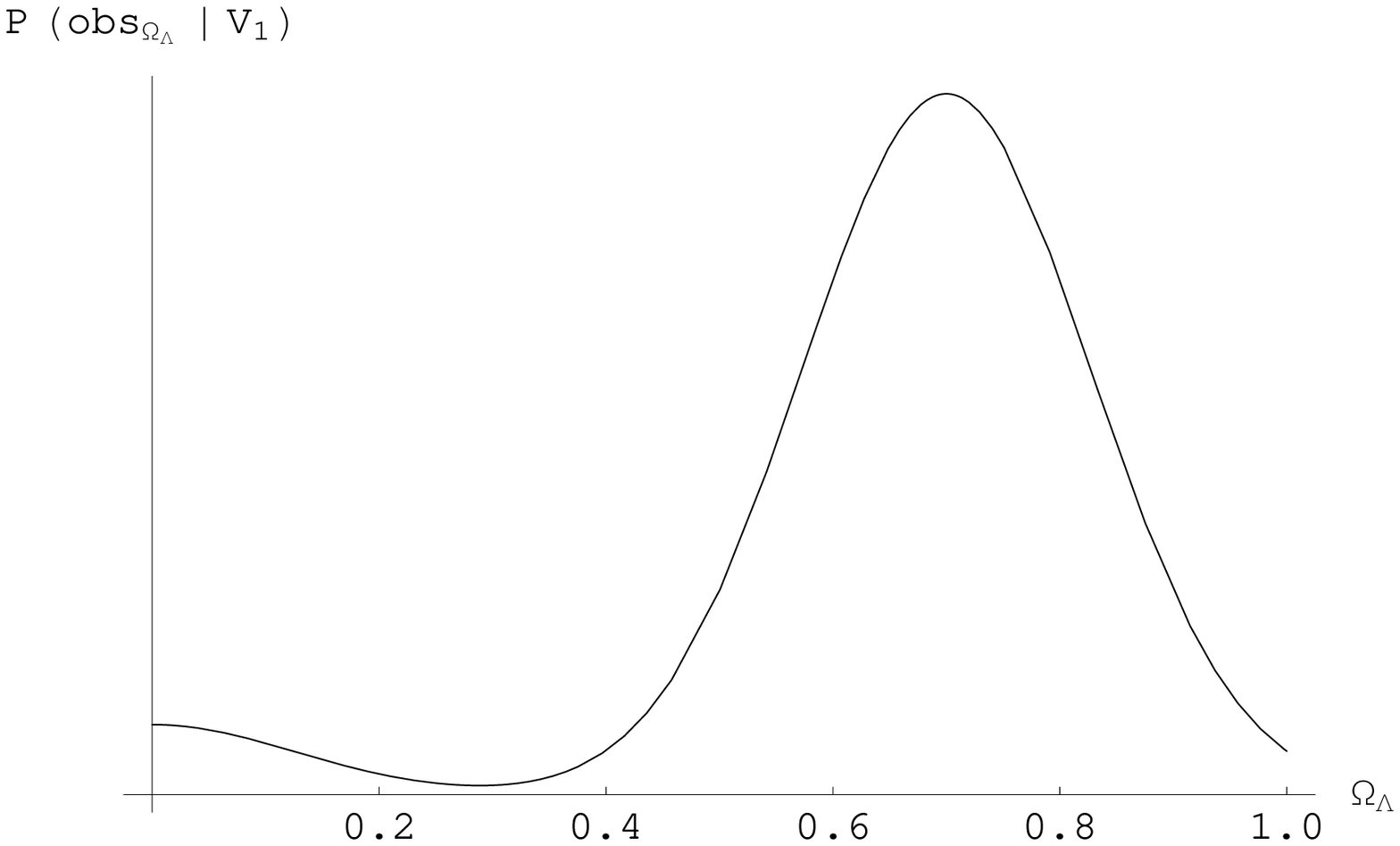,width=.6\textwidth}
\caption{The probability distribution for finding an observer who measures $\Omega_{\Lambda}=\rho_{\Lambda}/\rho_{\textrm{crit}}$ for a potential $V_1$. Since we measure $\Omega_{\Lambda}\equiv0.7$, we are a \emph{typical} observer given this potential $V_1$.}
\label{fig:totalprobabilitygood}}

\FIGURE[t]{\epsfig{file=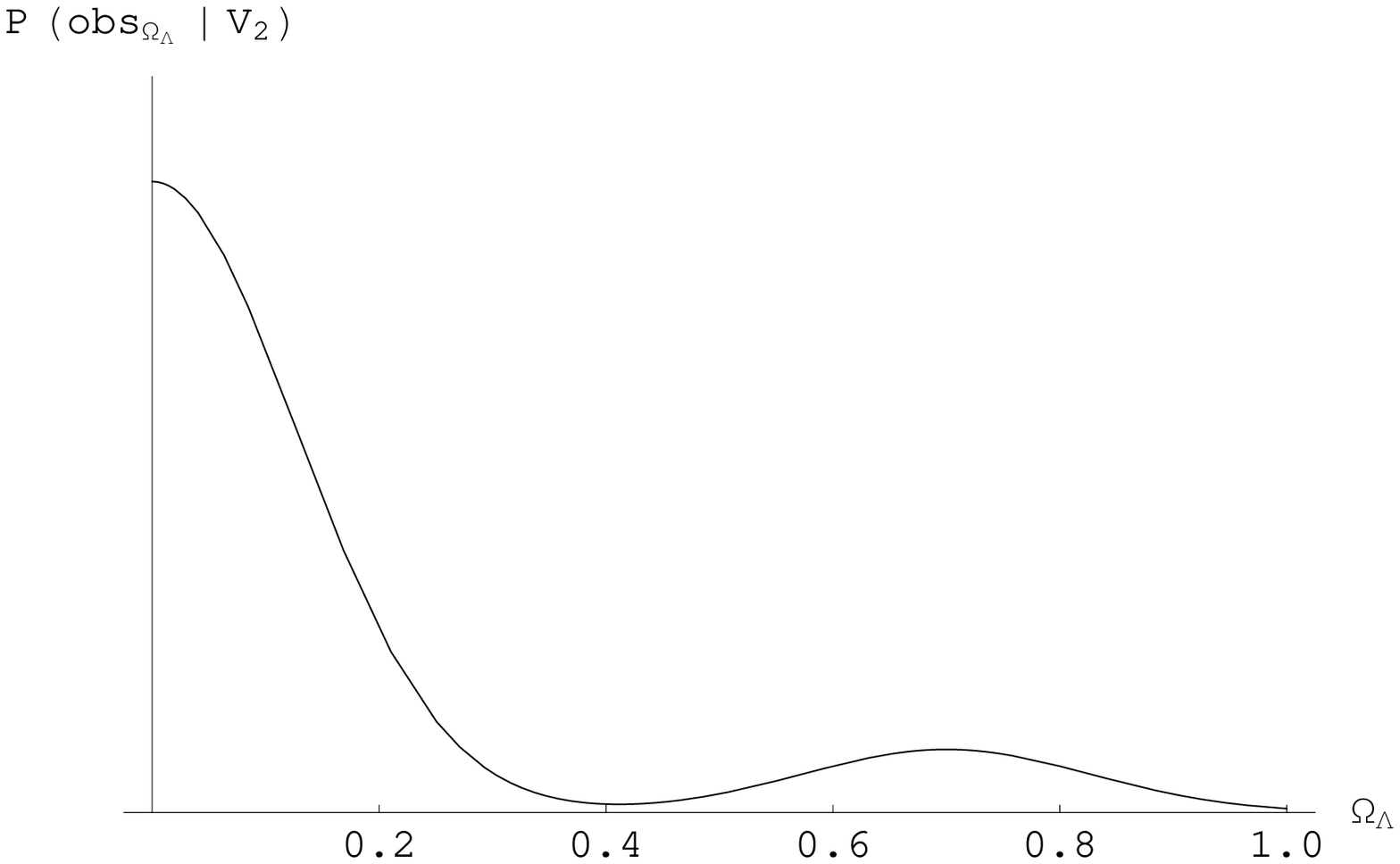,width=.6\textwidth}
\caption{The probability distribution for finding an observer who measures $\Omega_{\Lambda}=\rho_{\Lambda}/\rho_{\textrm{crit}}$ for a potential $V_2$. Since we measure $\Omega_{\Lambda}\equiv0.7$, we are an \emph{atypical} observer given this potential $V_2$.}
\label{fig:totalprobabilitybad}}

Let us be more specific about what we mean by ``typical'' and ``atypical''~\cite{Garriga:1998px}. The fact that there is only one observable universe is an extreme case of the cosmic variance problem that plagues observations of the cosmic microwave background (CMB) temperature anisotropy spectrum at low multipole moments~\cite{White:1993jr}. At each multipole $l$, there are only $2l+1$ independent measurements of the sky, resulting in an unremovable uncertainty in our knowledge of the universe compared to our \emph{expectation} of what the universe should look like on the basis of a given fundamental theory. In particular, these anisotropies are sourced by quantum fluctuations which are stochastic in nature, very much like the process of pocket universe generation via eternal inflation. Thus to unambiguously measure the \emph{distribution} from which the values for the CMB anisotropies are drawn would require observations over an ensemble of full-skies.

Similarly, to gain complete knowledge of the distribution of pockets (and hence possibly allowing us to derive the generating potential) one would need to measure an ensemble of pocket universes. The problem of cosmic variance is now even worse for many fundamental parameters than it is for the CMB.  For example, we are allowed only a single measurement for the case of $\Omega_{\Lambda}$. However accurately we measure this value, we cannot escape from the extreme cosmic variance that prevents us from learning the underlying distribution of $\Omega_{\Lambda}$. Again, typicalness is a matter of subjective interpretation; one can (arbitrarily) define typical as having more than $1$ in $1000$ odds\footnote{This definition means that our measurement of the CMB power at low multipole, which deviates from the expected value by about $1$ in a $100$, implies that we are a typical observer.}.  However, it is clear that given two different potentials, there is an intuitive notion of which is better: that potential for which we are more typical.

Finally, we hark back to our example and say that if a third potential $V_3$ generates the distributions such that  $P(1|V_3)\gg P(0|V_3),~P(\textrm{us}|V_3)$, then by distribution~(\ref{lambdadist}) the most common pocket has no support.  Hence we drop this pocket type from consideration and compare $P(0|V_3)$ with $P(\textrm{us}|V_3)$.  This point illustrates the utility of the $P(\textrm{obs}|\alpha_i)$, in that it allows us to eliminate the region of parameter space in which no observers can exist.

\section{Volume Based Approaches} \label{sec:volbased}

Before we discuss our method for computing $P(A|V)$, let us digress for a moment and review previous work on volume based approaches for comparison. Perhaps the most straightforward method for computing probability distributions arising from eternal inflation is to compare volumes. That is, we calculate the product of the first two probability factors, $\sum_A P\left(A|V(\psi_j)\right) P\left(\alpha_i |A\right)$, by considering the volume of the universe with parameters $\alpha_i$ divided by the total volume of the non-inflating portion of the multiverse. Since eternal inflation generates infinitely many pocket universes, to make this statement meaningful we must regulate the resulting infinite volumes in some way.  A volume based probability then looks like
\begin{equation}
P_{Vol}\left(\alpha_i|V(\psi_j)\right) = \lim_{x \to 0} \, \frac{Vol_x(\alpha_i)}{Vol_x(\textrm{all space})},
\end{equation}
where $x$ is a regulating parameter which is removed as $x \to 0$.  The difficulty arises from trying to perform the regularization in a gauge invariant manner.

To illustrate this difficulty, consider an eternally inflating space that is not past-infinite~\cite{Borde:2001nh} at a fixed value of a chosen time parameter.  This space is finite.  We may calculate any probabilities of interest and then take the time parameter to infinity to remove the regulator.  However, this method depends sensitively on which time parameter we choose.  With different choices of time, we could include more or less of some pocket universes and thereby alter the volume ratio.  In fact, since the pocket universes are space-like separated we may even choose some time parameter which excludes some pockets until arbitrarily late times.  Clearly, with a carefully chosen time parameter we can find nearly any answer we want~\cite{Guth:2000ka,Linde:1994gy,Winitzki:2005ya,Linde:1995uf}.

This sensitivity to the chosen time parameter is well illustrated by the so-called youngness paradox~\cite{Guth:2000ka} and by speculations that we live at the center of a large void~\cite{Linde:1994gy}, where in both cases the authors acknowledge that their results are crucially dependent on their choice of time.  In fact, Winitzki~\cite{Winitzki:2005ya} has shown that some gauge choices result in inflationary spaces which have decreasing volumes as time increases.  This counter-intuitive result shows that a na\"ive description of eternal inflation which uses the unbounded growth of volume is in fact not gauge invariant.  Fortunately, a gauge invariant definition of eternal inflation can be formulated on the basis of whether or not inflating regions exist at arbitrarily late times.

However, the use of volumes to define the probabilities is appealing and there have been attempts to do so without employing a time truncation.  One such idea is the spherical cut-off prescription~\cite{Vilenkin:1998kr,Vanchurin:1999iv} which is applicable if we have an inflationary potential with only one vacuum.  In that case, all of the pocket universes will be statistically identical, allowing us to consider only one pocket and ignore the others.  In this single pocket universe though, parameters may still vary and we can use volumes to determine the probabilities.  In this paper we have multiple different pocket universes, so this method provides an algorithm for computing the second probability factor, $P(\alpha_i|A)$.

To consider this method let us start by understanding the thermalization surface in the eternally inflating universe.  Suppose that by the time slow-roll inflation is ending there is one dominant degree of freedom: $\phi$.  The thermalization surface is thus a surface of constant $\phi$ where inflation ends. The end of inflation is understood via the semi-classical evolution of $\phi$, so we treat this surface as being continuous. On this surface, the time derivatives of the inflaton are necessarily much larger than the spatial gradients.  As a result, the gradient of the inflaton is a time-like vector,
\begin{equation}
g^{\mu\nu}\nabla_\mu \nabla_\nu \phi = \dot{\phi}^2 - \frac{1}{a^2}\left|\vec{\nabla}\phi\right|^2 > 0.
\end{equation}
The thermalization surface is therefore a space-like surface. If inflation is eternal then any thermalized region arbitrarily far in the future is still surrounded by an inflating region.  The surface thus extends to the infinite future.  However, since it is a space-like surface, we can find a coordinate system in which it appears spatially infinite (and nearly homogeneous).

Since volumes in single pocket universes are infinite we need to regularize them.  The spherical cut-off method does this by using a sphere of radius $R$ centered around a random point on the thermalization surface.  This sphere may be much larger than the horizon size.  Parameters may vary from one point to another within this sphere, and we calculate the probability for the parameters to take on any given value at an arbitrary point by using the volume inside this sphere.  Finally, we take the large $R$ limit.  By considering sufficiently large spheres we expect to have a representative sample of this pocket universe and the probability should converge.  Therefore the probability distribution for the parameters $\alpha_i$ given some pocket universe $A$ is
\begin{equation}
P\left(\alpha_i | A\right) = \lim_{R \to \infty} \frac{Vol_R(\alpha_i)}{4 \pi R^3 / 3}.
\end{equation}

In any context where the potential $V$ has many vacua, such as the string landscape, we also need to calculate the first factor in the probability distribution, $P(A|V)$. Previous attempts to compute this factor include using regenerating observers~\cite{Garriga:2001ri} or using a cut-off on the comoving size of pockets~\cite{Garriga:2005av}.  The first idea, unfortunately, requires arbitrary choices.  In this method imaginary observers are placed in the eternally inflating space at a constant rate per unit volume when the inflaton is near a local maximum of its potential.  The future location of the observers is then used to calculate probabilities for different thermalization vacua.  However, the relative rate of observer generation at different maxima is arbitrary.  Additionally, there is no unique definition of ``near a maximum''  to determine where observers should be generated.  Finally, this method results in an infinite number of observers to be counted.  It is therefore still not a regularization scheme for determining probabilities.  The second idea based on the comoving size of pockets appears to give the same results as our method described below.  This agreement is explained at the end of~\cite{Garriga:2005av}.  It remains to be seen which method will be more practical from a numerical point of view.  In the following section, we propose a different regulator, based not on volume, but on a finite set of world lines.

\section{World Line Method}\label{sec:ObserverMethod}

\subsection{The Main Idea: Sampling the Distribution} \label{subsect:mainidea}

In this section we describe our method of counting pockets. While our goal is to obtain the probability distribution of pockets generated by some potential $V$, the exact mechanism for eternal inflation is not important.  Indeed, for our example below we use a simple toy model to generate a spatial distribution of pockets.  Our pocket counting method is independent of this toy model and any other model which generates a spatial distribution of pockets is sufficient.  Our method will then sample these pockets by randomly drawing from this set a large number of times to find the probability distribution.

These random drawings are executed by laying down a finite set of $N$ points with some velocities on a compact patch in the initial hyper-surface, and then following the world lines from these points until they each reach some thermalized region of spacetime.  After throwing away some of the lines to avoid over-counting some pockets, we use the remaining lines as a representative sample of the pockets to calculate the probabilities. 

\subsection{Generating the Pockets: a Toy Model} \label{subsect:toymodel}

To generate our multiverse, we turn first to a common toy picture of eternal inflation~\cite{Guth:2000ka,Winitzki:2005ya}.  Consider inflation with a single scalar field and a single final vacuum state, given by the potential in Figure~\ref{fig:SingleVacuum}. We start with a small homogeneous inflating region which is taken to be the size of the Horizon, $1/H$.  In one discrete time step, which is not necessarily the Hubble time, the region expands and becomes some number, $N_H$, of new Hubble regions.  Due to the expansion, each new region has the same physical size as the original.  We take these new regions to be completely independent, and the fate of these new regions depends on the details of our model.  In the simplest case some fraction of these new regions thermalize and the remainder continue to inflate as before.  The same expansion/subdivision procedure is now applied to each of the new inflating regions.

\FIGURE[t]{\epsfig{file=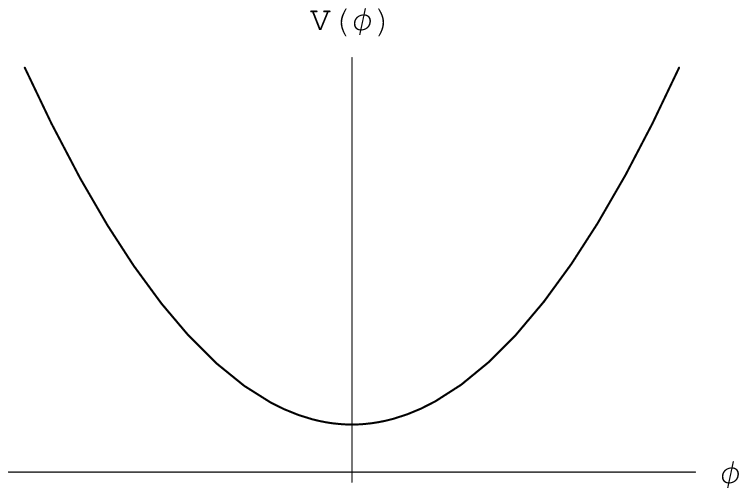,width=.6\textwidth}
\caption{A single vacuum symmetric potential with eternal inflation at $\phi > M_{Planck}$.  The inflaton rolls towards $0$ and thermalizes at the bottom.}
\label{fig:SingleVacuum}}

So long as there is at least one region which continues to inflate after the subdivision, inflation will be eternal.  If more than one region continues to inflate then the total volume of inflating regions grows in this gauge.  To be specific, suppose that a fraction, $q < 1$, of the new regions after subdivision continue to inflate.  Then picking a point in the initial space at random, we can calculate the probability to find this point remaining in an inflating region after $n_t$ time steps.  Despite the fact that the total inflating volume is growing, this probability goes to zero as $q^{n_t}$.  This is because at each time step, the probability to remain in an inflating region is $q$ and once the point has entered a thermalized region it will never experience inflation again.

Now, consider the slightly more complicated model where the inflaton can roll into either of two possible vacua\footnote{We assume that the tunneling time scale between the two vacua is sufficiently long.}, A or B, as in Figure~\ref{fig:NearlySymmetricPotential}.  Such an asymmetric potential would be realized with odd powers of $\phi$.  At the top of the hill, eternal inflation occurs.  At each time step the inflaton can thermalize to the left, thermalize to the right, or stay at the top and keep inflating. The probabilities for each outcome are $p_A$, $p_B$ and $q$ such that $1=q+p_A+p_B$. This model of eternal inflation generates type $A$ and type $B$ pocket universes.  Their ratio will be a function of the shape of the potential which is encoded in the probabilities.

\FIGURE[t]{\epsfig{file=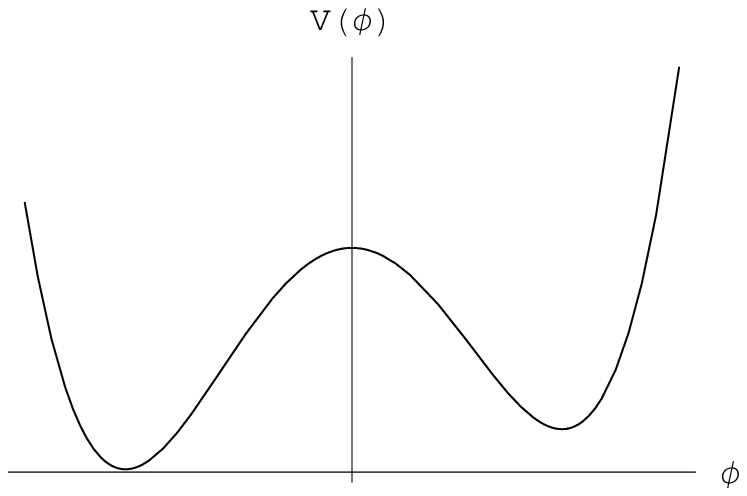,width=.6\textwidth}
\caption{A nearly symmetric potential with eternal inflation happening at the local max at $\phi=0$.  The inflaton may roll into and thermalize in either of the two vacua.}
\label{fig:NearlySymmetricPotential}}

This toy model of eternal inflation effectively assumes an explicit spatial foliation so that gauge problems are not transparent. However, we emphasize that our world line method is independent of the method for generating the multiverse.  We need only be given the spacetime which results from eternal inflation, and then our method will calculate the probability distribution of the pockets, $P(A|V(\psi_j))$.

\subsection{Counting the Pockets} \label{subsect:counting}

We start with a collection of $N$ points distributed randomly over the initial inflating region. We draw the location of the points from a flat probability distribution; as we will show later, our conclusions are not sensitive to the initial distribution of points or velocities. The probability to find one or more points still in any inflating region after $n_t$ time steps is
\begin{equation}
1- \left(1-q^{n_t}\right)^N,
\end{equation}
which, for large times, approaches zero as $N q^{n_t}$.  For any given set of $N$ points we run our ``pocket universe generation'' mechanism of the previous subsection long enough so that all of the points end up in thermal regions~\cite{Gratton:2005bi}. 

Our goal is to obtain the sample distribution of the pocket universes, not of the world lines.  Therefore if two or more world lines end up in the same pocket, we must throw out all but one to avoid over-counting that pocket.   Let $N'$ be the total remaining number of points after removing duplicates and $N_A$ be the number of the remaining points which end up in separate pockets of type A, then we conclude that the probability of generating vacuum A through eternal inflation is~\footnote{There exist probability distributions for other quantities which are convergent, such as the probability that inflation in a pocket lasted $n_e$ e-foldings, $P(A|n_e)$.  We leave these tangential computations for future work as they do not answer the question we set out in Section~(\ref{sec:ProbParts}).}
\begin{equation}
P(A | V(\psi_j)) = \frac{N_A}{N'}. \label{probequation}
\end{equation}

To see why we must throw out duplicates, consider an explicit example which results from the potential in Figure~\ref{fig:PathologicalPotential}. Eternal inflation will take place at the local maximum, which is exactly symmetric in this model.  The inflaton then has an equal chance of rolling either to the left or to the right.  If it rolls to the right it thermalizes in vacuum $B$.  However, if it rolls to the left, there will be another period of eternal inflation on the plateau where more pocket universes are created.  By construction, the inflaton may not fluctuate from the plateau back up to the maximum.

If the initial points are chosen randomly, then by symmetry, half of them roll to the right and half to the left.  If all of the points are kept, then we conclude that the probability of finding either vacuum is $1/2$.  However, the plateau to the left of the maximum allows for the production of many more pocket universes which thermalize in vacuum A.  In fact for each vacuum of type B which is created, there are an infinite number of vacua of type A.   If we remove duplicate world lines that end in the same thermalized region, the probability of finding vacuum A converges to one.

There is one more step we must add to this algorithm.  As eternal inflation progresses many more vacua are created at later times.  If the ratio of the number of vacua of type A and B changes as a function of time then we must somehow count the vacua at arbitrarily late times.  This is accomplished by taking the limit as the number of points grows without bound, $N' \to \infty$.  If the probability distribution converges, we have a useful calculational procedure. If the distribution does not converge then either the question is ill-posed or the underlying distribution of pockets is not well behaved.

\subsection{A Sample Calculation} \label{subsect:sample}
In this subsection, we begin with a slight generalization of the potential in Figure~\ref{fig:PathologicalPotential}; we allow the inflaton to hop back up from the plateau to the maximum via quantum fluctuation.  A related model is discussed in~\cite{Garriga:2005av}.  We list out the probabilities for each possible event in table~\ref{table:ToyProbabilities}.  In the limit that $p_1$ goes to zero and the maximum is symmetric, $p_2 = 1-q_2-p_2$, we recover the model discussed in the previous subsection.

\FIGURE[t]{\epsfig{file=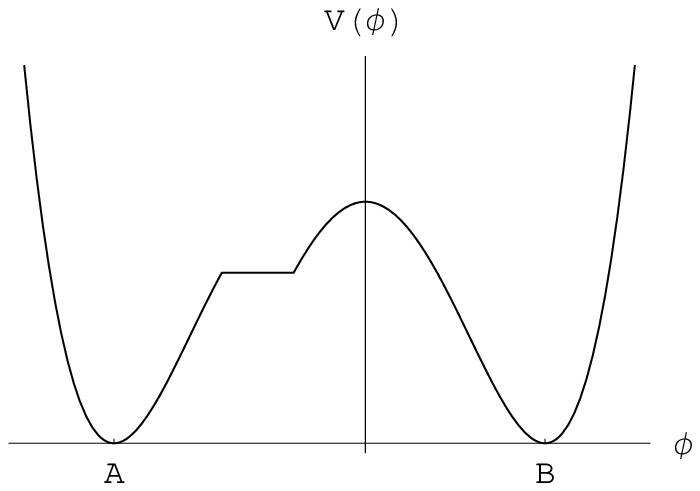,width=.6\textwidth}
\caption{An inflaton potential which is exactly symmetric at the local maximum, but which supports substantially more eternal inflation before rolling into vacuum A than vacuum B.}
\label{fig:PathologicalPotential}}

\TABULAR[t]{|c|l|}
{\hline 
\tt {\bf Probability} & {\bf Event} \\
\hline 
$q_1=0.5$ & Continue to inflate on plateau\\
$p_1=0.1$ & Hop from plateau up to maximum \\
$1-q_1-p_1$ & Drop from plateau and thermalize in vacuum A\\
\hline
$q_2=0.3$ & Continue to inflate at maximum\\
$p_2=0.4$ & Hop from maximum down to plateau\\
$1-q_2-p_2$ & Drop from maximum and thermalize in vacuum B\\
\hline}
{\label{table:ToyProbabilities}The assignment of probabilities for the toy example.}

More generally, it is straightforward, but tedious to calculate the probability of a comoving point to be in either vacuum A or B after $n_t$ time steps.  Let us suppose that the point starts in a region which is inflating at the maximum.  After one time step there is a probability $P_{max}=q_2$ that it will still be at the maximum and a probability $P_{pla}=p_2$ that it will be on the plateau.  A recursive expression gives the probabilities after $n_t$ time steps:
\begin{eqnarray}
P_{max}(n_t) &=& q_2 P_{max}(n_t-1) + q_1 P_{pla}(n_t-1) \nonumber \\
P_{pla}(n_t) &=& p_2 P_{max}(n_t-1) + q_1P_{pla}(n_t-1).
\end{eqnarray}
Since the inflaton cannot hop from the vacuum back up to the inflating regions in this model, we have monotonically increasing probabilities for the comoving point of interest to be found in the vacua:
\begin{eqnarray}
P_A(n_t) &=& P_A(n_t-1) + (1-q_1-p_1)P_{pla}(n_t-1) \nonumber \\
P_B(n_t) &=& P_B(n_t-1) + (1-q_2-p_2)P_{max}(n_t-1). \label{eqn:recursiveprob}
\end{eqnarray}
The inflaton at the chosen point must be in one of these four locations,
\begin{equation}
1 = P_A(n_t) + P_{pla}(n_t) + P_{max}(n_t) + P_B(n_t). \label{eqn:conservprob}
\end{equation}
Inserting Equations (\ref{eqn:recursiveprob}) into Equation (\ref{eqn:conservprob}), we find that the probability for the point to remain in an inflating region at each time step decreases by a positive fraction of the current probability to be in an inflating region:
\begin{equation}
-\Delta(P_{pla}+P_{max}) = (1-q_1-p_1)P_{pla} + (1-q_2-p_2)P_{max}. \label{eqn:convergence}
\end{equation}
We then see that, as in the simple example, the probability for any given point to continue to be in an inflating region goes to zero at late times.  From this fact that $P_{pla} \to 0$ and $P_{max} \to 0$, we can use Equation~(\ref{eqn:recursiveprob}) to see that the probabilities for a point to be in either vacuum are also converging.

We divide the space into $4$ new Hubble regions at each time step and choose the values listed in the table for the probabilities.  After iterating this expansion and subdivision procedure, we plot the number of regions of each type as a function of time step in Figures~\ref{fig:ToyModelInflatingCount} and~\ref{fig:ToyModelVacuumCount}.  Not surprisingly, at any given time step there are more regions inflating on the plateau than at the maximum and as a consequence more pocket universes in vacuum A are created.  Our intuition at this point would be that we should just stop the procedure and count the total number of pockets of each type.  If we did we would find that about $80\%$ of the pockets are of type A and $20\%$ of the pockets are of type B.  Of course, simply counting all pockets cannot be extended beyond the toy model in a gauge invariant manner.

\DOUBLEFIGURE[t]{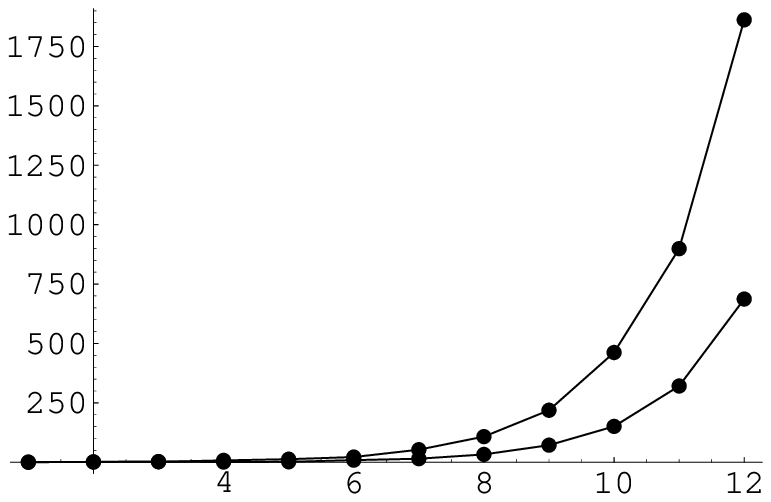,width=.4\textwidth}{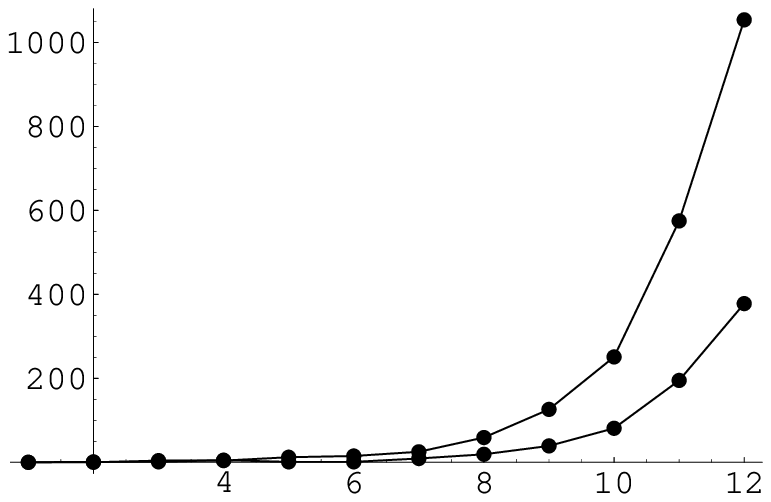,width=.4\textwidth}
{The number of the inflating regions as a function of time step.  The upper curve is the number of regions on the plateau.\label{fig:ToyModelInflatingCount}}
{The number of each type of vacua created as a function of time step.  The upper curve is the number of regions which fall in to vacuum A at that time step.\label{fig:ToyModelVacuumCount}}

We then choose random locations in the original space and check to see which pockets those points end up in.  This corresponds to following the world lines from each initial point when we extend this beyond the toy model setting.  We throw out multiple points in the same pocket and calculate the probability of finding either vacuum.  As we increase the number of points chosen, we check to see that the probability is converging.  The result is plotted in Figures~\ref{fig:ToyModelProbability1} and~\ref{fig:ToyModelProbability2}; the probability seems to be converging to about $80\%$.  From the statement below Equation~(\ref{eqn:convergence}) we know that this probability must converge, but we expect a variance of order $1/\sqrt{N'}$ from statistical fluctuations in picking the world lines.  This size of variation is consistent with the spread of the points in the figures.
\DOUBLEFIGURE[t]{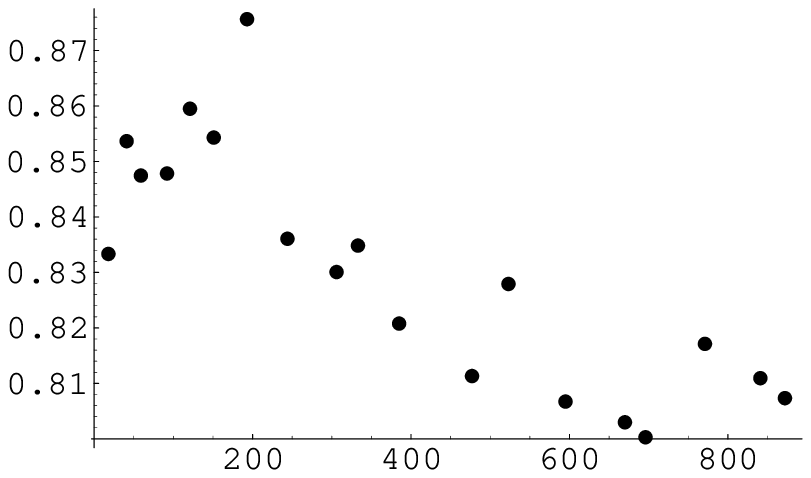,width=.4\textwidth}{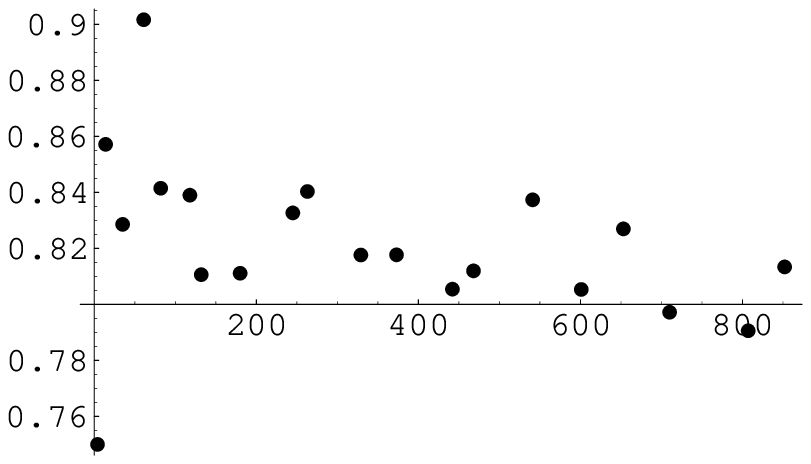,width=.4\textwidth}{The probability of vacuum A as a function of the number of non-duplicate points chosen.  The result seems to be converging to about $80\%$ as expected.  The starting locations are chosen from a uniform initial distribution.\label{fig:ToyModelProbability1}}
{Same as the previous Figure except the initial points are chosen from a distribution which is skewed towards one side of the space.\label{fig:ToyModelProbability2}}
The agreement between our two methods of calculating the probabilities in the toy model is reassuring and this latter method can be extended to a more realistic model of an eternally inflating spacetime.

If we had chosen to set $p_1=0$, we would find that the number of vacua of type A greatly exceeds that of type B and the probability of finding vacuum type A converges to one.  If we make the two inflating regions identical the probability converges to $1/2$.

\subsection{Insensitivity to the Initial Distribution of Points} \label{subsect:insensitive}
Although a different choice of initial distribution might result in variations to the rate at which the probability distributions converge, the final result is not sensitive to the initial distribution of points. This is because we remove duplicate world lines which fall into the same pocket universe.  If we compare two distributions, one of which is heavily weighted in one area, then more world lines from that area will end up in one or a few pocket universes. However, by eliminating these duplicates, both distributions give the same result. 

Our procedure could depend on the initial distribution if we have some prior knowledge of how the space will evolve when the points are placed. However, as long as we choose a distribution which is not informed by the future evolution of the space and take the large $N'$ limit, the result will be insensitive to the initial distribution. To illustrate this, we tested our toy model with two different distributions of initial points.  In the first case, we chose points from a uniform distribution on the initial slice.  In the second case, we used a distribution which was falling off as $1/\sqrt{x}$, where $x$ is some coordinate on the initial slice. The results are shown in Figures~\ref{fig:ToyModelProbability1} and~\ref{fig:ToyModelProbability2}.  Both cases converge toward the same result.  There are differences in the plots of order $1/\sqrt{N'}$, which is to be expected from the randomness of picking the world lines.

Similarly, any distribution of initial velocities for the world lines is equally good so long as the distribution is not designed to carefully pick out only a few pockets.

\section{Summary} \label{SummarySection} 
We have proposed a new method for calculating the probability distribution of different types of pocket universes in eternal inflation.  Because this method only involves following a finite number of world lines through an eternally inflating spacetime, the resulting probability is independent of the choice of gauge.  By discarding additional world lines which terminate in the same pocket, we make the final probability insensitive to the initial distribution of world lines.  Finally, we check for convergence of the probability as the number of world lines is increased.  This method avoids any reference to volume and only makes use of measurements along a set of world lines which are manifestly covariant.

In addition, we note that despite the extreme cosmic variance of having measurements from only a single pocket universe out of infinitely many, there exists a notion of how typical our observations are given a potential. If we assume that life is not sensitive to some parameters, such as small values of the cosmological constant, then we can observationally favor some inflationary models over others. This fact may ultimately demonstrate the utility of eternal inflation by allowing us to judge the viability of inflationary models.

\acknowledgments
R.E. and E.A.L. are supported in part by the DOE under contract DOE-FC02-92ER-40704. 
M.R.M. is supported in part by the DOE under contract W-7405-ENG-36 and would like to thank W. Goldberger for his hospitality when this work was begun. E.A.L. would like to thank Los Alamos National Laboratory for its hospitality where some of this work was done.  We would like to thank Cristian Armendariz-Picon, Gary Felder, Salman Habib, Jerome Martin and Alexander Vilenkin for useful and insightful discussions.

\end{document}